\documentclass[epsf,prb,twocolumn,showpacs]{revtex4-1}
\usepackage[pdftex]{graphicx}
\usepackage{dcolumn}
\usepackage{bm}
\usepackage{epsfig}
\usepackage{latexsym}
\usepackage{amsmath}
\usepackage{amsfonts}
\usepackage{amssymb}
\usepackage{color}
\usepackage{array}
\usepackage{framed}
\usepackage{esvect}

\renewcommand{\v}[1]{\mathbf{#1}} 

\newcommand{\T}{\mathcal{T}}

\newcommand{\be}{\begin{equation}}
\newcommand{\ee}{\end{equation}}
\newcommand{\bea}{\begin{eqnarray}}
\newcommand{\eea}{\end{eqnarray}}

\newcommand{\la}{\langle}
\newcommand{\ra}{\rangle}

\newcommand{\Z}{\mathbb{Z}}
\renewcommand{\vec}[1]{{\bf #1}}

\begin{document}

\title{Interacting fermionic topological insulators/superconductors in 3D}
\author{Chong Wang and T. Senthil}
\affiliation{Department of Physics, Massachusetts Institute of Technology,
Cambridge, MA 02139, USA}
\date{\today}
\begin{abstract}
 Symmetry Protected Topological (SPT) phases are a minimal generalization of the concept of topological insulators to interacting systems.  In this paper we describe  the classification and properties of such phases for three dimensional(3D)  electronic systems with a number of different symmetries.  {For symmetries representative of all classes} in the famous 10-fold way  of free fermion topological insulators/superconductors, we determine the stability to interactions. By combining with results on {\em bosonic} SPT phases we obtain a classification of   electronic 3D SPT phases for these symmetries. In cases with a normal $U(1)$ subgroup we show that this classification is complete.  We describe the non-trivial surface and bulk properties of these states. In particular we discuss interesting correlated surface states that are not captured in a free fermion description. 
{We show that in many, but not all cases, the surface can be gapped while preserving symmetry if it develops intrinsic topological order.}

\end{abstract}
\maketitle

\tableofcontents

\section{Introduction}

Much of our current understanding of topological insulators/superconductors is informed by models of free fermions and their associated band structure\cite{TIs}. Within this description there is a very mature understanding of the possible such phases in diverse dimensions. A classification of these free fermion topological phases exists\cite{tenfold} yielding results that depend on the global symmetry and the spatial dimensionality.   A defining characteristic of such phases is the presence of non-trivial surface states that are protected by the global symmetry. 

The free fermion description is clearly the appropriate starting point to discuss the possibility of topological insulators/superconductors in weakly correlated materials. In recent years however attention has turned toward materials with strong electron correlations as possible platforms for similar phenomena. These include the mixed valence compound\cite{tki} $SmB_6$, and iridium oxides on pyrochlore lattices\cite{irreview}. 

The exploration of topological phenomena in correlated materials brings with it a number of questions. Are the free fermion topological phases stable to the inclusion of interactions? 
Are there generalizations of topological insulators that have no free fermion analog? More generally how is the classification of topological phases of free fermion systems changed in a strongly interacting system? Clearly in addressing these questions there is a need to go beyond the concept of topological band structure and think more generally about the phenomenon of topological insulation.

Right at the outset  is important to note that there are many possible generalizations of the concept of topological insulator to interacting systems. An exotic possibility is to phases with a bulk gap which have `intrinsic' topological order\cite{Wenbook}. The classic examples are the fractional quantum Hall liquids and  gapped quantum spin liquids 
states of frustrated magnets. Intrinsically topologically ordered phases have multiple ground states on topologically non-trivial manifolds, and in the presence of symmetry may exhibit excitations with fractionalization of quantum numbers.  In contrast, topological band insulators/superconductors do not have intrinsic topological order.

In this paper we are concerned instead with a {\em minimal} generalization - known as Symmetry Protected Topological (SPT) phases - of the free fermion topological phases to interacting systems.  These phases have a  bulk gap and no intrinsic topological order but nevertheless have interesting surface states that are protected by global symmetries.  The classic example of a Symmetry Protected Topological phase is the Haldane spin-$1$ chain. This has a bulk gap, no fractionalization, and non-trivial end states that are protected by symmetry. In $d = 1$ all SPT phases have been classified\cite{1dsptclass} in the last few years.

For systems of interacting bosons (or the closely related interacting spin systems), there is by now a good understanding of the possibility and physics of SPT phases in all physical dimensions ($d = 1, 2, 3$).  This progress was initiated by a formal classification\cite{chencoho2011} - based on the concept of group cohomology - of bosonic SPT phases. Though this classification is now known not to be complete\cite{avts12,hmodl,ashvinbcoh,kapustin} (in three dimensions), it represents substantial progress.   The physical properties of various such bosonic SPT phases are not simply obtained through the cohomology classification. These have been described by other physics-based methods in both 
two\cite{levingu,luashvin,tsml,liuxgw}  and in three dimensions\cite{avts12,xuts13,hmodl,ashvinbcoh,metlitski,avdomain}.

In contrast to bosonic systems, our understanding of fermionic SPT phases beyond band theory is rather limited, particularly in the physically important case of three space dimensions.  An interesting attempt\cite{scoho}  to generalize the formal cohomology method to fermions lead to a concept known as group super-cohomology and to some results on fermion SPT phases. However currently this formal method is not able to handle the physically important cases of either continuous symmetry or the Kramers structure of the electron.  
In dimension $d = 2$ however a simpler Chern-Simons approach provides many definitive results for fermionic SPT states\cite{luashvin}. The effect of strong interaction was also examined for certain kinds of $2D$ fermionic SPT states described by band theory\cite{2dfermions}, where it was found that some of the topological bands became trivial in the presence of strong interactions.  {These approaches, however, are} difficult to generalize to higher dimensions.

In a recent paper\cite{3dfSPT} we (together with A. Potter) classified and described the physical properties of such interacting three dimensional electronic topological insulators in the physically important situation where both charge conservation and time reversal symmetries are present.  The $Z_2$ classification of such insulators within band theory was shown to be modified to a $Z_2^3$ classification in interacting systems, resulting in a total of 8 distinct phases. These are generated by 3 `root' states of which one is the topological band insulator and the other two are Mott insulators where the spins form a spin-SPT phase (various models of such ``topological paramagnets'' were discussed in Ref. \onlinecite{chencoho2011, hmodl, ashvinbcoh}).  The physics-based methods used in Ref. \onlinecite{3dfSPT}  enabled us  to obtain a very clear picture of the physical properties of the various states and determine their experimental fingerprints. 
It was also shown there that insulators without time-reversal symmetry ($U(1)$ only) have no non-trivial SPT phase.

 {In this paper we generalize the ideas of Ref. \onlinecite{3dfSPT} to discuss $3d$ electronic topological insulators/superconductors with many other symmetries. Free fermion  topological phases with various symmetries fall into one of 10 distinct classes. This is known as the 10-fold way\cite{tenfold}. With interactions there is no  guarantee that systems with two different symmetries that fall in the same class in the 10-fold way still have the same possible SPT phases. Therefore it is important to specify the symmetry group directly. For symmetry groups  represented in each of the famous 10-fold way we are able to ascertain the stability of the free fermion classification to interactions.} If the symmetry group has a normal $U(1)$ subgroup we obtain a complete classification of the interacting electronic SPT states.  The results are summarized in Table. \ref{classtable}.  

For time reversal invariant superconductors in three dimensions (class $DIII$) a recent paper\cite{TScSTO} showed that the $Z$ classification of band theory reduces to a $Z_{16}$ classification with interactions. For this symmetry we provide a simpler derivation of the same result. For other symmetry classes our results have not been described in the literature as far as we know.

 \begin{table*}[tttt]
\begin{tabular}{|>{\centering\arraybackslash}m{1.5in}|>{\centering\arraybackslash}m{1.2in}|>{\centering\arraybackslash}m{1in}|>{\centering\arraybackslash}m{1in}|}
\hline
{\bf Symmetry class} &  {\bf Reduction of free fermion states} & {\bf Distinct boson SPT} & {\bf Complete classification} \\ \hline
$U(1)$ only (A) & $0$ & $0$ & $0$ \\ \hline
$U(1)\rtimes\Z_2^T$ with $\T^2=-1$ (AII) & $\Z_2\to\Z_2$ & $\Z_2^2$ & $\Z_2^3$ \\ \hline
$U(1)\rtimes\Z_2^T$ with $\T^2=1$ (AI) & $0$ & $\Z_2^2$ & $\Z_2^2$ \\ \hline
\hline
$U(1)\times\Z_2^T$ (AIII) & $\Z\to\Z_8$ & $\Z_2$ & $\Z_8\times\Z_2$ \\ \hline
$U(1)\rtimes(\Z_2^T\times\Z_2^C)$ (CII) &  $\Z_2\to\Z_2$ & $\Z_2^4$ & $\Z_2^5$ \\ \hline
$(U(1)\rtimes\Z_2^{T})\times SU(2)$ & 0 & $\Z_2^4$ & $\Z_2^4$ \\ \hline
\hline
$\Z_2^T$ with $\T^2=-1$ (DIII) & $\Z\to\Z_{16}$ & $0$ & $Z_{16}$ (?) \\ \hline
$SU(2)\times\Z_2^T$ (CI) & $\Z\to\Z_4$ & $\Z_2$ & $Z_4 \times Z_2$ (?) \\ \hline

\end{tabular}
\caption{Summary of results on classifications of electronic SPT states in three dimensions. The second column gives free fermion states that remain nontrivial after introducing interactions. The third column gives SPT states that are absent in the free fermion picture, but are equivalent to those emerged from bosonic objects such as electron spins and cooper pairs. For symmetries containing a normal $U(1)$ subgroup, we can find the complete classification. In all such examples the complete classifications are simple products of those descending from free fermions and those obtained from bosons. For symmetry class $C$I, we give suggestive arguments but not a proof that the classification in the last column is complete. 
}
\label{classtable}
\end{table*}%

\section{Generalities}
\label{general}

It is useful to first describe a few general ideas that will form the basis of the physical arguments used to establish our results. 

\subsection{Surface terminations}
 
A crucial property of an SPT phase is the presence of non-trivial surface states protected by the global symmetry. It is thus no surprise that powerful constraints are obtained by thinking about the possible surface terminations of the bulk SPT phase, {\em i.e} different possible surface phases that correspond to the same bulk phase. 
The surface either spontaneously breaks the symmetry, or if gapped, has intrinsic topological order. A gapless symmetry preserving surface is also in principle possible. More fundamentally any effective theory for the surface implements symmetry in a manner not possible in a strictly two dimensional theory.

 \subsubsection{ Symmetry broken surface}

A particularly useful surface termination is one where the defining global symmetry is either partially or completely broken. In the latter case the surface can be fully gapped 
without introducing intrinsic topological order. This follows from the assumption that the phase is symmetry protected. The non-triviality of the symmetry broken surface  manifests itself in the topological defects of the symmetry breaking order parameter.  This ensures that we cannot produce a trivial symmetry preserving surface by proliferating topological defects.  

We mention two particularly interesting examples of broken symmetries here. The first one is the breaking of time-reversal symmetry, which can be realized explicitly by depositing a ferromagnet on the surface. Very often (but not always) the domain walls between opposite $\mathcal{T}$-breaking regions hosts chiral modes, which prohibits the domain walls to proliferate and restore $\mathcal{T}$. The chiral modes in the domain wall are related to quantized Hall conductance (say, of charge, spin or heat)  in each of the domains. We will discuss Hall transport in more detail in Sec.\ref{thetaterm}.

The second example is a surface that breaks $U(1)$ symmetry.  If the $U(1)$ symmetry corresponds to charge-conservation, this can be realized by depositing a superconductor on the surface. Below we will use the terminology of superconductivity to describe the $U(1)$ symmetry breaking more generally (even if the $U(1)$ symmetry does not actually correspond to charge conservation).  It is well known that the $U(1)$ symmetry can be restored by proliferating (condensing) vortices. Therefore if the ``superconductor" is gapped (and has no intrinsic topological order), the fundamental ($hc/2e$) vortex must be non-trivial.  Otherwise it can be proliferated to restore a trivial insulator on the surface. However, there always exist some higher vortices that are trivial in terms of statistics and symmetry representation, and thus can be condensed. In this case a topologically ordered surface arises, which will be discussed further in Sec.\ref{spsto} and throughout the paper.

\subsubsection{Symmetry preserving surface topological order}
\label{spsto}

Powerful insights into the SPT phase are provided by a surface termination which is fully gapped and preserves the symmetry at the price of having intrinsic topological order just at the surface. This was first demonstrated in the context of bosonic SPT phases\cite{avts12}. Conceptually such a  topologically ordered surface state provides a nice and non-perturbative characterization of the bulk SPT order\cite{avts12,hmodl,metlitski,ashvinbcoh,TScSTO,fSTO1,fSTO2,fSTO3,fSTO4}. We point out here that it is not always guaranteed that such a symmetry preserving surface topological ordered phase will exist. Indeed later in the paper we will discuss an example where a symmetry preserving surface is necessarily gapless. When a symmetric surface topologically ordered state exists, it too must realize symmetry in a manner forbidden in strictly two dimensional systems. 

\subsection{Gauging the symmetry: $\theta$ terms}
\label{thetaterm}
Another useful theoretical device is to formally gauge all or part of the defining global symmetry to produce a new physical system. This can be done for all unitary symmetries or for unitary subgroups of the full symmetry group.  Two cases will be of particular interest to us. In the first case the full symmetry group $G$ has a normal $U(1)$ subgroup which we can then consistently gauge while retaining the quotient group $G/U(1)$ as an unbroken global symmetry. In the second case the continuous part of the full global symmetry is  $SU(2)$. In this case there is no normal $U(1)$ subgroup and instead we gauge the full continuous $SU(2)$ symmetry. 

Let us first discuss the case where there is a normal $U(1)$ subgroup to which we couple a gauge field. As the bulk is gapped we may formally integrate out the electrons and obtain an effective  {long wavelength} Lagrangian for the gauge field. 
\begin{equation}
{\cal L}_{eff} = {\cal L}_{Max} + {\cal L}_\theta
\end{equation}
The first term is the usual Maxwell term and the second is the `theta' term:
\begin{equation}
{\cal L}_\theta = \frac{\theta}{4\pi^2} \v{E}\cdot\v{B}
\end{equation}
where $\v{E}$ and $\v{B}$ are the electric and magnetic fields respectively of the $U(1)$ gauge field.  The allowed value of $\theta$ will be constrained by the unbroken 
global symmetry $G/U(1)$. In the familiar example of the topological band insulator (with $G = U(1) \rtimes Z_2^T$, where $Z_2^T$ is time reversal), it is well known that $\theta = 0$ or $\pi$ by time reversal symmetry. This is true as well for other symmetry groups in Table. \ref{classtable} that include time reversal. The $\v{E}$ and $\v{B}$ fields transform oppositely under $Z_2^T$ so that $\theta \rightarrow - \theta$. 
Further on a closed manifold there is periodicity under $\theta \rightarrow \theta + 2\pi$ so that the only distinct possibilities are $\theta = 0, \pi$. 

The $\theta$ term provides very useful constraints on the surface physics. It can be written as the derivative of a Chern-Simons term. Hence at a surface where the $Z_2^T$ symmetry is broken, it leads to a  Hall conductivity (associated with transport of the $U(1)$ charge) of 
\begin{equation}
\sigma_{xy} \equiv \nu =  \frac{\theta}{2\pi}
\end{equation}
(We use units in which the $U(1)$ charge of the fermions is $1$ and $\hbar = 1$).
Furthermore  in all the examples studied in this paper, such a $Z_2^T$ broken surface termination exists with a gap and without any surface topological order. In that case we can safely say that when $\nu$ is fractional , the surface state cannot exist in strictly two dimensional systems, and requires the $3d$ bulk. Thus fractional $\nu = \frac{\theta}{2\pi}$ for the response to a $U(1)$ gauge field implies non-trivial bulk SPT order even in the presence of interactions. 

Returning to the $Z_2^T$ broken gapped surface without any topological order, we can further argue that the {\em difference} in the Hall conductivity of the surface and its time reverse must be a state allowed in $2d$ systems of electrons without topological order. This forces $\frac{\theta}{\pi} = n$ with $n$ an integer. 

Another important characterization of such a $Z_2^T$ broken surface state is the thermal Hall conductivity $\kappa_{xy}$.  Formally this is  related to gravitational responses in the bulk and the notion of gravitational anomaly\cite{ganomaly} though we will not need to use such a description.  For any gapped  two dimensional state $\nu_Q = \frac{\kappa_{xy}}{\kappa_0}$ is a universal number with $\kappa_0  = \frac{\pi^2}{3}\frac{k_B^2 }{h} T$ ($T$ is the temperature).  For a {\em strictly} two dimensional system if further there is no topological order then $\nu_Q - \nu = 0~~ (mod ~8)$ (see Appendix. \ref{e8}). Thus a gapped $Z_2^T$ broken surface that either has fractional $\nu$ or has 
$\nu_Q - \nu \neq 0~~ (mod ~8)$ implies non-trivial bulk SPT order even with interactions. 

In the case where the continuous symmetry is $SU(2)$ (e.g., associated with spin conservation) we can again gauge this $SU(2)$ symmetry and study the effective Lagrangian of the corresponding matrix-valued $SU(2)$ gauge fields $A_\mu$  which again takes the form 
\begin{equation}
{\cal L}_{eff} = {\cal L}_{Max} + {\cal L}_\theta
\end{equation}
The first Maxwell term is the usual Lagrangian for the $SU(2)$ gauge field ($g$ is a non-universal coupling constant). 
\begin{equation}
{\cal L}_{Max} = \frac{1}{2g} Tr\left( F_{\mu \nu} F_{\mu \nu} \right)
\end{equation}
The field strength $F_{\mu \nu} = \partial_\mu A_\nu - \partial_\nu A_\mu + [A_\mu, A_\nu]$. The second `theta' term takes the form
\begin{equation}
{\cal L}_\theta = \frac{\theta}{32 \pi^2} Tr \left(\epsilon_{\mu\nu\alpha\beta} F_{\mu\nu} F_{\alpha\beta}\right)
\end{equation}
On a closed manifold there is periodicity under $\theta \rightarrow \theta + 2\pi$. Time reversal if present takes $\theta \rightarrow -\theta$, and thus the potentially time reversal invariant possibilities are $\theta = n\pi$. 
This $\theta$ term can once again be written as the derivative of a Chern-Simons term for the $SU(2)$ gauge field: 
\begin{equation}
{\cal L}_\theta = \frac{\theta}{8\pi^2} \partial_\mu K_\mu
\end{equation}
where 
\begin{equation}
K_\mu = \epsilon_{\mu\nu\alpha \beta} Tr\left( A_\nu \partial_\alpha A_\beta + \frac{2}{3} A_\nu A_\alpha A_\beta \right)
\end{equation}

Similarly to the discussion of the $U(1)$ case above this implies that a $Z_2^T$ broken gapped surface without topological order will have a {\em spin quantum Hall effect}  \cite{tsbmmf99}. This is characterized by the spin current induced in the transverse direction in response to a  spatially varying Zeeman field. (The spin quantum Hall effect should not be confused with the quantum spin Hall effect - the latter describes the transverse spin current induced by an electrical voltage). The corresponding spin Hall conductivity $\sigma^s_{xy} = \frac{\theta}{\pi}$. Note the factor of $2$ difference between the corresponding formula for the $U(1)$ case. In a strictly $2d$ system we must have $\sigma^s_{xy} = 2n$ with $n$ an integer. Therefore an odd $\frac{\theta}{\pi}$ implies bulk SPT order 
even with interactions. 

In both $U(1)$ and $SU(2)$ cases, if the $\theta$ term is such that the $Z_2^T$ broken surface state has Hall transport that is allowed in $2d$ we cannot directly conclude anything about whether an SPT state exists or not. In the following subsection we obtain some additional constraints in the $U(1)$ case by thinking about monopole defects of the gauge field.

\subsection{Gauging the symmetry: bulk monopoles and surface states}

An important lesson from the work in Ref. \onlinecite{levingu} is that when the global symmetry in an SPT phase is gauged the defects of the gauge field could become nontrivial. 
Let us now consider the situation discussed above where the global symmetry has a normal $U(1)$ subgroup which we then gauge. 
Then the gauge defect is simply the magnetic monopole: a $2\pi$-source of the gauge flux. In three dimensions the monopole statistics can only be bosonic or fermionic. It was shown in Ref. \onlinecite{3dfSPT} that in any (short-range entangled)  system where all the charge-$1$ particles are fermions, the monopole must be a boson.  The monopole may then carry non-trivial quantum numbers under the symmetry group.

The `electric' charge of the monopole under the $U(1)$ symmetry is determined directly by the $\theta$ term in the effective gauge action through the well-known Witten effect: there is a a $U(1)$ charge of $\frac{\theta}{2\pi}$ on the monopole. For $\theta = \pi$ this is fractional. The remaining question is about the symmetry transformation of the monopole under the quotient group $G/U(1)$. In particular it will be important to ask of the monopole transforms under a projective representation of this quotient group. 
In the familiar case of electrons with $U(1) \rtimes Z_2^T$, the symmetry properties of the monopole under $Z_2^T$ are severely constrained\cite{3dfSPT}. The monopole goes to an antimonopole under $Z_2^T$ and this makes it meaningless to ask about whether time reversal acts projectively on it or not.  In the $U(1)\times\Z_2^T$ case studied in detail below, the gauge magnetic flux is even under time-reversal.  Hence it is also possible to have monopoles forming nontrivial (projective) representations under time-reversal, {\em i.e} it could become a Kramers doublet, with $\T^2=-1$. It is possible to enumerate all possible nontrivial quantum numbers that can be carried by the monopole. For example, with $U(1)\times\Z_2^T$, the monopole can either carry half-integer $U(1)$ charge (corresponding to $\theta= \pi$), or  be charge-neutral while having $\T^2=-1$, or both. 

Understanding the allowed structure of the bulk monopole leads to important constraints on the possible surface terminations of the SPT.  Such a point of view was nicely elaborated in Ref. \onlinecite{metlitski} to discuss the physics of the bosonic topological insulator. We emphasize that this procedure of gauging the symmetry and studying the monopole is a purely theoretical device. It is however very powerful. 

It is convenient for our purposes in this paper to consider a surface termination which breaks the $U(1)$ symmetry. Imagine tunneling a monopole from the vacuum into the system bulk, which leaves behind a two-fold ($hc/e$ using the `superconducting' terminology) vortex on the surface. As the monopole is trivial in the vacuum, if it carries nontrivial quantum numbers in the bulk, the corresponding $hc/e$ vortex on the surface must also carry the same nontrivial quantum numbers, and vice versa. Therefore we could either use the known monopole property to infer the properties of the surface `superconductor' (as was done in Ref. \onlinecite{fSTO1}), or use the knowledge of the surface vortex to infer the quantum numbers carried by the bulk monopole (as was done in Ref. \onlinecite{hmodl}, and will be done in Sec.\ref{AIII} and \ref{CII} below).

It is important to emphasize that not all the seemingly consistent projective symmetry representations of monopoles can actually be realized. For example, in a spinless fermion system where $\T^2=1$ on the fermions, a theory with half-charged monopole ($\theta=\pi$) is naively consistent, even though there is no free fermion band structure that realizes such a theory. One may wonder if there is an intrinsically interacting SPT state with no free fermion realization that gives $\theta=\pi$. However, it was noticed recently\cite{3dfSPT,fSTO2} that it is internally inconsistent in a subtle way, and hence cannot be realized even with an interacting SPT. Therefore if a symmetry assignment of the monopole is not realized in any free fermion system, one needs to examine its consistency more carefully. 

The next question is then, if a symmetry assignment to the monopole is realized by a representative state (for example, a free fermion model), how many other states exist with the same monopole properties? For such a state with certain monopole quantum numbers, there is always a representative state with monopoles carrying the ``opposite'' quantum numbers, so that stacking the two states together produces another state with monopoles carrying trivial quantum numbers. Therefore the question can be posed equivalently as: how many nontrivial states exist with monopoles being completely trivial? This was analysed in detail in  Ref. \onlinecite{3dfSPT}, and for completeness we give  {the argument} in Appendix. \ref{tm}. The conclusion is that if an SPT state has trivial monopole, it must be equivalent to a SPT state constructed  from bosonic particles carrying no $U(1)$ charge (e.g. spins in an electron system). For example, if a fermionic SPT state with $U(1)\rtimes\Z_2^T$ symmetry has a trivial monopole, it is equivalent to a bosonic SPT with $\Z_2^T$ symmetry only.

The above conclusion can be summarized compactly as follows:

\begin{framed}
 If the group $G$ contains a normal $U(1)$ subgroup, then any 3D SPT state with the symmetry group $G$ must either have a nontrivial $U(1)$ monopole (one that transforms projectively under $G$), or be equivalent to a bosonic SPT with the symmetry $G/U(1)$.
\end{framed}

In the rest of the paper we utilize these different ways of diagnosing and differentiating SPT phases to classify and understand electronic SPT phases with many different symmetries.

\section{$U(1)\times\mathbb{Z}_2^T$: AIII class}
\label{AIII}

In this section we study fermions with the symmetry group $U(1)\times \mathbb{Z}_2^T$ (the AIII class), which can be interpreted physically as superconductors with $S_z$-spin conservation and time-reversal symmetry. The $U(1)$ rotation $U(\theta)$ and time-reversal $\mathcal{T}$ commutes: $U(\theta)\mathcal{T}=\mathcal{T}U(\theta)$, or equivalently, the $U(1)$ charge is odd under time-reversal action, unlike the electric charge. Physically, the action of time-reversal on the fermions has two distinct possibilities: $\mathcal{T}^2c\mathcal{T}^{-2}=\pm c$, where $c$ is the physical fermion annihilation operator. However, the two symmetries lead to very similar physics, including the classification of SPT states. This is because one can always define a new time-reversal-like operation $\tilde{\mathcal{T}}=\mathcal{T}U(\pi/2)$, and it is easy to see that $\tilde{\mathcal{T}}^2=-\mathcal{T}^2$ on the fermion. Hence the problem with $U(1)\times\mathbb{Z}_2^T$ with $\mathcal{T}^2=-1$ on the fermion can be mapped to that with the same symmetry group $U(1)\times\mathbb{Z}_2^{\tilde{T}}$ but with $\tilde{\mathcal{T}}^2=1$. We will take $\mathcal{T}^2=-1$ below in order to be able to connect to other interesting symmetry groups, but the modified time-reversal $\tilde{\mathcal{T}}$ will still be useful as a tool in our argument.

The free fermion band theory gives a $\mathbb{Z}$ classification for this symmetry group. Each state is labeled by an integer $n$ signifying the number of protected gapless Dirac cones on the surface:
\be
\label{dirac}
H=\sum_{i=1}^n\psi_i^{\dagger}(p_x\sigma_x+p_y\sigma_z)\psi_i,
\ee
with the symmetries acting as $U(\theta): \psi\to e^{i\theta}\psi$ and $\mathcal{T}: \psi\to i\sigma_y\psi^{\dagger}$.

We will show in the following that the $\mathbb{Z}$ classification from band theory reduces to $\mathbb{Z}_8$ in the presence of interaction: the $n=1$ state has a bulk $\theta=\pi$ term, the $n=2$ state, which has $\theta=2\pi$, has a neutral Kramers monopole ($\mathcal{T}^2=-1$), the $n=4$ state is equivalent to an SPT state formed by bosons carrying no $U(1)$ charge, hence the $n=8$ state, formed by taking two copies of the $n=4$ state, is trivial. Following the arguments in Sec.\ref{general}, we can also show that taking another bosonic SPT (which cannot be realized using free fermions) into account, we obtain the complete classification given by $\mathbb{Z}_8\times\mathbb{Z}_2$.

\subsection{8 Dirac cones: triviality}
\label{8cones}

We first look at the $n=8$ state, which has eight protected surface Dirac cones in the free fermion theory. We will show explicitly that, with interaction, such surface state can open up a gap and become a trivial state.  {We use an argument very similar to that in Ref. \onlinecite{3dfSPT} (see Supplementary Materials). }

We first introduce a singlet pairing term into the theory
\be
\label{pairing}
H_{\Delta}=\sum_{i=1}^{n}i\Delta\psi\sigma_y\psi+h.c.,
\ee 
which breaks both the $U(1)$ and $\mathcal{T}$ symmetries (under time-reversal we have $\Delta\to-\Delta^*$). The surface theory is now gapped, with the physical symmetries broken. With interactions, the gap $\Delta$ becomes a fluctuating field, hence it is possible to disorder it (have $\la\Delta\ra=0$) and restore the symmetries. To disorder the $XY$-like field $\Delta$, we can follow the familiar and well-understood route of proliferating vortices of the order parameter. 

It is important here to notice that although the gap in Eq.~\eqref{pairing} breaks both $U(1)$ and $\mathcal{T}$, it does preserve a time-reversal-like subgroup generated by $\tilde{\mathcal{T}}=\mathcal{T}U({\pi/2})$. Since we want to restore $\mathcal{T}$ by disordering $\Delta$ (which surely will restore $U(1)$), we must do it while preserving $\tilde{\mathcal{T}}$. 


The vortex needs to be examined carefully because of the fermion zero-modes associated with it. It is well-known that a superconducting Dirac cone gives a Majorana zero-mode in the vortex core\cite{fkmajorana}. Under $\tilde{\mathcal{T}}$, the vortex background is invariant (unlike the case of $U(1)\rtimes\Z_2^T$, where a vortex goes to an anti-vortex), and the Majorana zero-modes transform trivially $\gamma_i\to\gamma_i$. At free fermion level the degeneracy from the zero-modes are robust, since any quadratic term $iA_{ij}\gamma_i\gamma_j$ would break $\tilde{\mathcal{T}}$. However, it is known\cite{1dfSPT} that with $n=8$ Majorana zero-modes, the degeneracy can be lifted by a quartic term. The remaining vortex is then completely trivial and can thus be condensed without breaking $\tilde{\mathcal{T}}$, producing a trivial insulator on the surface.

We can also examine the time-reversal properties of the vortices more directly, which will be useful in the following sections. We first pair up the 8 Majorana zero-modes to 4 complex fermion zero-modes $f_i=\gamma_{2i-1}+i\gamma_{2i}$. We then define different vortex operators as
\begin{equation}
v_{nmkl}|GS\rangle=\left(f_1^{\dagger}\right)^n\left(f_2^{\dagger}\right)^m\left(f_3^{\dagger}\right)^k\left(f_4^{\dagger}\right)^l|FN\rangle,
\end{equation}
where $|FN\rangle$ denotes the state with all the negative-energy levels filled in a vortex background. The $U(1)$ being spin-like under $\mathcal{T}$ (hence $\tilde{\mathcal{T}}$ also) means that a vortex configuration is time-reversal invariant. The only non-trivial action of $\tilde{\mathcal{T}}$ is thus on the zero-modes:
\begin{equation}
\tilde{\mathcal{T}}f_i\tilde{\mathcal{T}}^{-1}=f^{\dagger}_i,
\end{equation}
and by choosing a proper phase definition:
\begin{equation}
\tilde{\mathcal{T}}|FN\rangle=f^{\dagger}_1f^{\dagger}_2f^{\dagger}_3f^{\dagger}_4|FN\rangle.
\end{equation}

It is then straightforward to check the modified time-reversal $\tilde{\mathcal{T}}$ only relates vortex operators with the same fermion parity  $(-1)^{n+m+k+l}$, and $\tilde{\mathcal{T}}^2=1$ on all the $v_{nmkl}$ operators. Moreover, vortices with the same fermion parity are mutually local with each other, and thus can be condensed simultaneously while keeping the $\tilde{\mathcal{T}}$ symmetry. The remaining surface is then a trivial gapped symmetric state.

\subsection{4 Dirac cones: boson SPT}
\label{4cones}

We now look at the state with $n=4$ Dirac cones on the surface and try to do the same exercise as in Sec.\ref{8cones}. Again we start from a paired gapped state and try to disorder the pairing gap by proliferating vortices. We have now $n=4$ Majorana zero-modes in the vortex core, and even with interaction the degeneracy cannot be lifted. The two-fold vortex ($hc/e$), on the other hand, hosts 8 Majorana zero-modes and hence is trivial. Condensing the two-fold vortex will then give a symmetric gapped state, with an intrinsic $\mathbb{Z}_2$ topological order\cite{z2long}, {\em i.e} described by a deconfined $Z_2$ gauge theory. To study this $\mathbb{Z}_2$ topologically ordered state, we need to examine the fundamental vortex in the superconducting state with more care.

Again we group the 4 Majorana zero-modes into two complex fermion zero-modes $f_{1,2}$, and define vortices through
\begin{equation}
\label{vortex}
v_{nm}|GS\rangle=\left(f_1^{\dagger}\right)^n\left(f_2^{\dagger}\right)^m|FN\rangle,
\end{equation}
where $|FN\rangle$ denotes the state with all the negative-energy levels filled in a vortex background.  The modified time-reversal again acts as
\begin{equation}
\label{zmtr}
\tilde{\mathcal{T}}f_{1,2}\tilde{\mathcal{T}}^{-1}=f^{\dagger}_{1,2},
\end{equation}
and by choosing a proper phase definition:
\begin{equation}
\label{fntr}
\tilde{\mathcal{T}}|FN\rangle=f^{\dagger}_1f^{\dagger}_2|FN\rangle.
\end{equation}
It then follows straightforwardly that $\{v_{00},v_{11}\}$ and $\{v_{01},v_{10}\}$ form two "Kramers" pairs under $\tilde{\mathcal{T}}$ (namely $\tilde{\mathcal{T}}^2=-1$). Moreover, since the two pairs carry opposite fermion parity, they actually see each other as mutual semions.

We thus conclude that to preserve the symmetry, the ``minimal" construction is to proliferate double vortices. The resulting insulating state has $\mathbb{Z}_2$ topological order $\{1,e,m,\epsilon\}$ with the $e$ being the remnant of $\{v_{00},v_{11}\}$, $m$ being the remnant of $\{v_{01},v_{10}\}$, and $\epsilon$ is the neutralized fermion $\tilde{\psi}$.

Now the full $U(1)\times\mathcal{T}$ is restored, and we can ask how they are implemented on $\{1,e,m,\epsilon\}$. Obviously these particles are charge-neutral, so the question is then about the implementation of $\mathcal{T}$ alone. However, since the particles are neutral the extra $U(1)$ rotation in $\tilde{\mathcal{T}}$ is irrelevant and they transform identically under $\tilde{\mathcal{T}}$ and $\mathcal{T}$. Hence we have $\mathcal{T}^2=\tilde{\mathcal{T}}^2=-1$ on $e$ and $m$, and $\mathcal{T}^2=\tilde{\mathcal{T}}^2=1$ on $\epsilon$. This state is denoted as $eTmT$ in  Ref. \onlinecite{hmodl}, and is a characteristic surface state of a bosonic SPT. We thus conclude that the $n=4$ free fermion state is equivalent to the $eTmT$ bosonic SPT in the presence of interaction.

\subsection{2 Dirac cones: Kramers monopole}

The $n=2$ state, being a "square root" of the $n=4$ state which is equivalent to a bosonic SPT, must involve the $U(1)$ symmetry in a non-trivial manner as argued in Sec.\ref{general}. It must thus have monopoles that are non-trivial under the symmetries. It turns out that the charge-neutral monopole behaves as a Kramers pair under time-reversal ($\mathcal{T}^2=-1$). Such monopole behavior was also realized in a boson SPT state\cite{hmodl} with $U(1)\times\Z_2^T$ symmetry, where charge-$1$ carriers are bosons instead of fermions. So in contrast with charge-$1/2$ monopoles ($\theta=\pi$), the Kramers monopole can be realized in two different systems, one with fermionic charge carriers and one with bosonic one. 

We show this by studying the monopole tunneling event on the surface: if the monopole has $\mathcal{T}^2=-1$ inside the insulator bulk and $\mathcal{T}^2=1$ in the vacuum outside, the tunneling event on the surface must leave behind another excitation with $\mathcal{T}^2=-1$. We can work this out directly from the free fermion surface state, by showing that a monopole insertion operator in a $(2+1)-d$ theory with two Dirac cones has $\mathcal{T}^2=-1$ due to the fermion zero-modes from the Dirac cones. An alternative route, which we will follow, is to study the paired state described in Sec.\ref{8cones} and \ref{4cones}, in which a monopole tunneling event leaves behind a two-fold ($hc/e$) vortex that now traps four Majorana zero-modes. The argument in Sec.\ref{4cones} immediately shows that this two-fold vortex has $\tilde{\T}^2=-1$, which means the monopole inside the bulk also has $\tilde{\T}^2=-1$. But since the monopole is charge-neutral, it has $\T^2=\tilde{\T}^2=-1$.

  \subsubsection{Surface topological order}
  \label{2conesTO}

The $n=2$ state can also be analyzed in a similar fashion as for $n=8$ and $n=4$ states. As we have noticed, the two-fold vortex in the paired state has $\tilde{\T}^2=-1$ and hence cannot be condensed to restore time-reversal symmetry. Hence the minimal construction is to condense the four-fold vortex, which traps eight Majorana zero-modes and can be trivially condensed. A charge-$1/2$ boson (denoted as $\beta$) emerges from this non-trivial vortex condensate, which under time-reversal goes to $\beta\to\beta^{-1}\sim\beta^3$ (the last identification comes from the topological triviality of $\beta^4$). The particle content of the remaining theory can be represented as $\{1,\beta,\beta^2,\beta^3, v,v,^2,v^3, \beta^nv^m\}\times\{1,c\}$, where v is the remnant of the fundamental vortex with the complex fermion zero-mode un-occupied, and $c$ is the physical fermion, while the remnant of the $\psi$ fermion is denoted as $\epsilon=c^{\dagger}\beta^2$.

The resulting gapped surface state has $\mathbb{Z}_4$ topological order, with the symmetries implemented in a peculiar way. The remnant of the fundamental vortex with the complex fermion zero-mode un-occupied is $v$, and that with the zero-mode occupied is $\epsilon v$. The two go to each other under time-reversal, and their squares (either $v^2$ or $\epsilon v^2$) have $\T^2=\tilde{\T}^2=-1$, with $\epsilon$ having $\T^2=1$. The topological sector of $\beta^2$ does not change under time-reversal, but since it carries charge-$1$, we have $\T^2=-\tilde{\T}^2=-1$ for it, which is consistent with $\T^2=1$ on $\epsilon$, since $\beta^2\epsilon\sim c$. The charge-vortex relation gives the obvious mutual statistics $\theta_{\beta v}=e^{i\pi/2}$.

\subsection{1 Dirac cone: $\theta=\pi$}

The $n=1$ state is the $U(1)\times\mathbb{Z}_2^T$ counterpart of the familiar electronic topological band insulator\cite{FKM}. The surface single Dirac cone implies a $\theta$-term in the gauge response\cite{qi} at $\theta=\pi$. The monopole then carries charge-$1/2$.

  \subsubsection{Non-Abelian Surface topological order}
  \label{1coneTO}
  
Following the reasoning from previous sections, we know that in the paired surface state, the four-fold ($2hc/e$) vortex is Kramers under $\tilde{\T}$ and we need to condense eight-fold vortex to recover the full symmetry. A charge-$1/4$ boson $\alpha$ emerges out of this condensate, and as in the $n=2$ case, the charge-$1$ boson $\alpha^4$ has $\T^2=-\tilde{\T}^2=-1$.
  
The story about lower vortices, however, is made more complicated due to the structure of the zero-modes. In particular, the fundamental vortex carries only one Majorana zero-mode and is thus non-abelian. The detailed analysis of the fusion and statistics of the vortices was carried out in  Ref. \onlinecite{fSTO1,fSTO2}, which showed that the fundamental vortex has topological spin $1$ while the two-fold vortices have topological spin $\pm i$, depending on whether the complex fermion zero-mode is filled or not. Fusing the vortex with an $\epsilon$ fermion gives back the vortex: $v\times\epsilon\sim v$, while fusing two vortices gives either the semionic or anti-semionic two-fold vortex: $v\times v\sim v^2+\epsilon v^2$. The mutual statistics between the fundamental and two-fold vortices are $\pm i$, hence the three-fold vortex has topological spin $-1$. It also follows that the fundamental vortex and the four-fold vortex (which is Kramers) are mutual semions (mutual statistics $-1$). 

Again the particle content can be written as $\{1,\alpha,...\alpha^7,v,...v^7, \alpha^nv^m\}\times\{1,c\}$.

We summarize the properties of different vortices in different states in Table.\ref{vortextable}.

 \begin{table}[tt]
\begin{tabular}{|>{\centering\arraybackslash}m{1.4in}|>{\centering\arraybackslash}m{1.2in}|}
\hline
{\bf Vortex zero-modes} &  {\bf Properties}  \\ \hline
$8$ Majorana & Trivial \\ \hline
$4$ Majorana & $\T^2=-1$ \\ \hline
$2$ Majorana, $2$-fold ($hc/e$) vortex & semion/anti-semion \\ \hline
$2$ Majorana, fundamental ($hc/2e$) vortex & bosonic, $\T: v\to \epsilon v$  \\ \hline
$1$ Majorana & Non-abelian \\ \hline
\end{tabular}
\caption{Summary of vortex properties, according to the number of Majorana zero-modes trapped. Most of the properties do not depend on the vortex strength (as long as the vortex exists), except when there are two Majorana zero-modes. In $n=1$ phase such a vortex has strength-$2$ while in $n=2$ phase it has strength-$1$, and the vortex statistics turns out to be different in the two cases. }
\label{vortextable}
\end{table}%

\subsection{$\mathbb{Z}_8\times\mathbb{Z}_2$ classification}
\label{82classification}

We have shown that the $\mathbb{Z}$ classification from free fermion band theory reduced to $\Z_8$ under interaction. The argument outlined in Sec.\ref{general} makes it possible to further classify all the SPT states, including those not realizable using free fermions.

For any putative new SPT phase that cannot be realized using free fermions, there is always a free fermion state such that the combination of the two has a trivial monopole. This is because every possible nontrivial symmetry implementation of the monopole is realized by a free fermion model. Following the reasonings in Sec.\ref{general}, a phase with trivial monopole can at most be a SPT made of charge-neutral bosons (with $\Z_2^T$ symmetry only). Bosonic SPT states with $\Z_2^T$ symmetry in three dimensions are classified by $\Z_2^2$, with two root states\cite{avts12}. One of the two root states becomes identical to the $n=4$ free fermion state. Hence it does not give rise to any new state. But the other root state is independent of all the free fermion states. Hence it provides a new state in the full classification. The final result is thus a $\Z_8\times\Z_2$ classification of three dimensional fermions with $U(1)\times\Z_2^T$ symmetry. 

\section{$\mathbb{Z}_2^T$ with $\mathcal{T}^2=-1$: DIII Class}

In this section we apply the results obtained in Sec.\ref{AIII} to superconductors with only time-reversal symmetry (the DIII class). This was recently discussed in Ref. \onlinecite{TScSTO}  {using powerful Walker-Wang methods}.   {We reproduce part of the results there in a physically simpler and constructive approach\cite{vrtxavetal} following the ideas of Ref. \onlinecite{3dfSPT} and the previous section}. 

At free fermion level, the DIII class superconductors in 3D are classified by $\Z$, with an integer index $\nu$ signifying the number of gapless Majorana cones on the surface protected by time-reversal symmetry:
\be
H=\sum_{i=1}^{\nu}\chi_i^{\dagger}(p_x\sigma_x+p_y\sigma_z)\chi_i.
\ee
If $\nu$ is even ($\nu=2n$), one can group the Majorana cones into $n$ Dirac cones $\psi_i=\chi_{2i-1}+i\chi_{2i}$, and the theory looks exactly the same as Eq.~\eqref{dirac}. The $U(1)$ symmetry $\psi\to e^{i\theta}\psi$ is now an emergent symmetry at low energy. We can instead consider the $U(1)$ as a microscopic symmetry, apply the results in Sec.\ref{AIII} to obtain interacting gapped surface states, and then break the $U(1)$ symmetry explicitly by adding fermion pairing term.   {A similar strategy was useful in the Walker-Wang approach\cite{TScSTO}}. For the $n=8$ ($\nu=16$) state, the resulting surface is trivially gapped, and further breaking the $U(1)$ symmetry does not introduce anything nontrivial. Hence the $\Z$ classification from band theory reduces to $\Z_{16}$ with interaction. For the $n=4$ ($\nu=8$) state, the resulting surface is topologically ordered, but all the quasi-particles are charge-neutral under the $U(1)$, hence breaking $U(1)$ symmetry does not affect anything either. These establish the $\nu=16$ state as a trivial one, and the $\nu=8$ state as equivalent to a boson SPT, which are consistent with the results in  Ref. \onlinecite{TScSTO}. The $n=2$ ($\nu=4$) and $n=1$ ($\nu=2$) states, however, have surface topological orders involving the $U(1)$ symmetry non-trivially, hence need more careful examination.

\subsection{4 Majorana cones: doubled semion-fermion surface state}

We now take the surface topological order in Sec.\ref{2conesTO}, break the $U(1)$ symmetry but keep time-reversal. Notice that $\T^2=-1$ on both $\beta^2$ and $v^2$, hence the simplest particle to condense is the charge-$1$ object $v^2\beta^2$. It can be checked straightforwardly that the remaining theory contains the following deconfined particles (and their combinations):
\begin{align}\{1,s_1=v\beta\}\times\{1,s_2=c v^{-1}\beta\}\times\{1,c\} \label{eq:dsemionfermion},\end{align} 
where $c$ is now the charge-neutral physical fermion. The mutual statistics between $\beta$ and $v$ in the original theory $\theta_{\beta,v}=i$ makes the composite $s_1=v\beta$ a semion with self-statistics $i$, likewise the particle $s_2=cv^{-1}\beta$ is also a semion. Under time reversal, $s_1=v\beta\to(v\epsilon)\beta^{-1}=s_1\epsilon\beta^2=s_1c$ which is an anti-semion, likewise $s_2\to s_2 c$ which is again an anti-semion. The two semions $s_1,s_2$ are local with respect to each other, and their bound state $s_1s_2=c\beta^2=\epsilon$ is a fermion with $\T^2=1$. These are in agreement with the result in  Ref. \onlinecite{TScSTO}.

\subsection{2 Majorana cones: semion-fermion surface state}

The fate of the surface topological order in Sec.\ref{1coneTO} is more complicated. Again $\T^2=-1$ on both $\alpha^4$ and $v^4$, and the simplest particle to condense is the charge-$1$ object $v^4\alpha^4$. It can be checked that the only effect of this condensate is to confine odd powers of $\alpha$: $\alpha^{2n+1}$. The remaining theory can then be written in the following way:
\begin{align}
\label{TPf}
\{1,v,...v^7\}\times\{1,s=\epsilon\alpha^{2}v^2\}\times\{1,c\},
\end{align}
where $\epsilon=c\alpha^4=cv^4$. It can be checked that particles in the two sectors $\{1,v,...v^7\}$ and $\{1,s\}$ are mutually local with respect to each other. The sector $\{1,v...v^7\}\times\{1,c\}$, with its time-reversal implementation, is exactly what was named T-Pfaffian state in  Ref. \onlinecite{fSTO3} and was proposed to be a possible surface state of the electronic band topological insulator\cite{fSTO3,fSTO4}. The only difference here is that there is no charge assignment. The T-Pfaffian state being a surface state of the band TI implies that without charge assignment (i.e. when charge $U(1)$ is broken), it should be possible to completely confine it down to $\{1,c\}$. This is a highly non-trivial statement, since there is no trivial boson in the theory for one to condense, and one need a series of unknown phase transitions to confine it. Now taking the statement as true, we can eliminate the $\{1,v...v^7\}$ sector from Eq.~\eqref{TPf} and get $\{1,s\}\times\{1,c\}$. Recall that $v^2$ is a semion, it also has $-1$ mutual statistics with $\alpha^2$ from the charge-vortex relation. Hence the composite $s=\epsilon\alpha^2v^2$ is a semion, and under time-reversal it goes to $s=\epsilon\alpha^2v^2\to \epsilon\alpha^{-2}\epsilon v^2=\alpha^{-2}v^2=(\alpha^{-4}\epsilon)s=cs$ which is an anti-semion. These are in agreement with  Ref. \onlinecite{TScSTO}.

\section{$SU(2)\times\mathbb{Z}_2^T$: CI Class}
\label{c1}

The results in Sec.\ref{AIII} can also be applied to systems with $SU(2)\times\Z_2^T$ symmetry, i.e. superconductors with full spin rotation and time-reversal symmetry: the CI class. Again the free fermion bands are classified by $\Z$. For a state indexed by $k$, there are $2k$ Dirac cones on the surface, giving $k$ flavors of $SU(2)$-fundamental fermions:
\be
\label{su(2)dirac}
H=\sum_{i=1}^k\psi_i^{\dagger}(p_x\sigma_x+p_y\sigma_z)\otimes\tau_0\psi_i,
\ee
where $\tau_{\mu}$ is the $SU(2)$ spin, so that the $SU(2)$ rotation $\mathcal{U}$ acts as
\be
\mathcal{U}: \psi_i\to\sigma_0\otimes \mathcal{U}\psi_i,
\ee
and time-reversal acts as
\be
\mathcal{T}:\psi_i\to i\sigma_y\otimes\tau_0\psi_i^{\dagger}.
\ee

 {At $k = 1$ when the surface is gapped by breaking time reversal, there is a spin quantum Hall effect of $\sigma^s_{xy} = 1$. This is half of what is allowed in $d = 2$. Correspondingly if we gauge the global $SU(2)$ symmetry the bulk response has a $\theta$ term\cite{ryuetalmodel} for the corresponding $SU(2)$ gauge field at $\theta = \pi$. As we argued earlier the $k = 1$ state is therefore stable to interactions. }

 {As in previous sections there is an emergent $U(1)$ symmetry in the surface Dirac theory}:
\be
U(\theta):\psi_i\to e^{i\theta}\psi_i,
\ee
which we can promote to a physical symmetry, apply the arguments in Sec.\ref{AIII} and get a gapped state, then break the $U(1)$ by an explicit pairing. One should, however, be careful in the procedure not to break the $SU(2)$ symmetry. It turns out for even $k$, it is possible to have an intermediate $U(1)$-breaking phase preserving the $SU(2)$ symmetry, while for odd $k$ this is impossible.  Hence the results from Sec.\ref{AIII} can be applied to $k=4$ (8 Dirac cones) to show that it is trivial, and to $k=2$ (4 Dirac cones) to show that it is equivalent to the $eTmT$ boson SPT. We show the latter in Sec.\ref{su(2)4cones} since it directly implies the former due to the $\Z_2$ nature of the corresponding boson SPT states. For the $k=1$ (2 Dirac cones) state, we argue in Sec.\ref{su(2)2cones} that it is impossible, even with interactions, to gap out the surface state while keeping the full $SU(2)\times \Z_2^T$ symmetry. Interestingly, it is so far the only known example in 3D with a symmetry protected gapless surface robust even under strong interaction.

The above results lead to a partial classification given by $\Z_4\times\Z_2$, where the $\Z_4$ subgroup was deduced from the $\Z$ classification in free fermions, and the $\Z_2$ subgroup comes from boson SPT states with $\Z_2^T$ symmetry, as discussed in Sec.\ref{82classification}. Unlike the symmetries with a normal $U(1)$ subgroup, it is not clear in this case if other SPT phases exist with no analog in either free fermion or boson systems. The analysis below in Sec. \ref{hopf} suggests (but does not prove) that non-trivial surface states beyond boson SPT can be described by a Hopf term in the non-linear-sigma model, which prevents the surface from opening up a trivial gap. Since the Hopf term is realized in a free-fermion model, this suggests that states beyond boson SPT are either free fermion phases, or the combination of boson SPT and free fermion phases, hence the above $\Z_4\times\Z_2$ classification may be complete. 
Likewise, superconductors with only $SU(2)$ symmetry may not support any nontrivial SPT state, since the surface Hopf-angle can always be tuned to zero in the absence of time-reversal symmetry. It is desirable to make the above arguments precise.

\subsection{4 Dirac cones: boson SPT}
\label{su(2)4cones}

We first re-write the $k=2$ surface Dirac state as
\be
H=\psi^{\dagger}(p_x\sigma_x+p_y\sigma_z)\otimes\tau_0\otimes\mu_0\psi,
\ee
where $\mu$ denotes the flavor index. We now write down the pairing gap term:
\be
H_{\Delta}=i\Delta\psi\sigma_y\otimes\tau_y\otimes\mu_y\psi+h.c.,
\ee
which obviously opens up a gap and preserves $SU(2)$ invariance. As in Sec.\ref{AIII}, time-reversal and the $U(1)$ symmetries are broken, but the modified time-reversal $\tilde{\T}=\T U(\pi/2)$ is kept invariant.

The vortex of $\Delta$ field carries four Majorana zero-modes, or two complex fermion zero-modes $f_{1,2}$. Since $SU(2)$ symmetry is kept and the $\psi$ fermion is an $SU(2)$-fundamental, the two complex fermion zero-modes must also form an $SU(2)$ doublet $(f_1,f_2)^T$. Again we define vortices through Eq.~\eqref{vortex}, and time reversal acts as in Eq.~\eqref{zmtr} and \eqref{fntr}. It is then clear that $\{v_{00},v_{11}\}$ are $SU(2)$ singlets and $\{v_{01},v_{10}\}$ form an $SU(2)$ doublet. Both pairs are Kramers under $\tilde{\mathcal{T}}$ ($\tilde{\mathcal{T}}^2=-1$). Moreover, since the two pairs carry opposite fermion parity, they actually see each other as mutual semions. Condensing two-fold vortices then gives the $\Z_2$ topological order $\{1,e,m,\epsilon\}$, where $e\sim\{v_{00},v_{11}\}$, $m\sim\{v_{01},v_{10}\}$ and $\epsilon\sim\psi$. All the particles are neutral under the $U(1)$, hence further breaking the $U(1)$ symmetry does not change anything in the topological order. Now both the $e$ and $\epsilon$ particles are $SU(2)$ doublets, so we can bind them with a physical fermion $c$ to produce $SU(2)$ singlets. The topological order can thus be re-written as $\{1,\tilde{e},m,\tilde{\epsilon}\}$, where $\tilde{e}=c\epsilon$ and $m$ have $\mathcal{T}^2=-1$, and $\tilde{\epsilon}=ce$ has $\T^2=1$, and all the particles are $SU(2)$ trivial. This is indeed the $eTmT$ state promised.

\subsection{2 Dirac cones: symmetry-enforced gaplessness}
\label{su(2)2cones}

With two Dirac cones one cannot write down a gap term that breaks $U(1)$ but not $SU(2)$, hence the previous trick does not apply. In fact, as we will now argue on very general grounds, it is impossible to have a gapped (topologically-ordered) symmetric surface state for the $k=1$ topological superconductor. Hence the two Dirac cones on the surface are robust even with strong interaction, as long as the full $SU(2)\times\Z_2^T$ symmetry is preserved.

If the surface can be symmetrically gapped by introducing a topological order, then the $SU(2)$ group has to be represented non-projectively for all the particles in the theory, since there is no projective representation for $SU(2)$. One can then always bind a non-trivial quasi-particle with certain number of physical fermions to form an $SU(2)$ singlet. Therefore the theory can always be re-written as $\{1,\epsilon,....\}\times\{1,c\}$ where all the particles are $SU(2)$ singlets except $c$. The first sector is also closed under time-reversal, since time-reversal action cannot mix an $SU(2)$ doublet with a singlet. Any local object in the topological order $\{1,\epsilon,...\}$ must then be bosonic since it is $SU(2)$ trivial. Hence the topological order can be viewed as one emergent from bosonic objects in the theory, and the bulk state can at most be a bosonic SPT state with $\Z^{T}_2$ symmetry only.

For the $k=2$ state the above analysis is consistent with what we obtained in Sec.\ref{su(2)4cones}. The $k=1$ state, on the other hand, cannot fit into the above framework: putting two copies of the $k=1$ state together forms a $k=2$ state, which is a bosonic SPT. The bosonic SPT's in this case are classified by $\Z_2^3$, so none of them admits a "square root". So the $k=1$ state cannot be a bosonic SPT, and according to the above analysis we are forced to conclude that a symmetric gapped surface topological order does not exist for the this state. This provide the first known example of "strictly" symmetry-protected gapless surface, since all the other 3D SPT states studied so far admit a gapped symmetric surface with topological order. This also implies that the $k=1$ (and the combination of the state with other boson SPTs) cannot be constructed using the Walker-Wang approach\cite{ashvinbcoh,TScSTO,fSTO3}, which relies on the existence of a gapped surface.

 \subsubsection{$O(3)$ non-linear sigma model: Hopf term}
 \label{hopf}

The surface Dirac theory in Eq.~\eqref{su(2)dirac} can also be gapped by introducing a Neel-like order. For $k=1$ we write down the Dirac fermion coupled to the Neel unit vector $\vec{n}$:
\be
H=\psi^{\dagger}(p_x\sigma_x+p_y\sigma_z)\otimes\tau_0\psi+m\psi^{\dagger}\sigma_y\otimes \vec{n}\cdot\vec{\tau}\psi.
\ee
Since the fermion is gapped now, one can integrate it out and obtain an effective theory of the Neel vector. The result\cite{abanovwiegmann} is a non-linear sigma model with a topological term known as the Hopf term, at $\theta=\pi$:
\be
S=\frac{1}{g}\int d^2xdt(\partial_{\mu}\vec{n})^2+i\pi H_2[\vec{n}],
\ee
where $H_2$ is the integer characterizing $\pi_3(S^2)=\Z$.

The Hopf term changes the statistics of the skyrmions of the $O(3)$ model\cite{fwaz}. Continuum field theory arguments suggest that time reversal (and parity) are preserved so long as the coefficient of the Hopf term is $0$ or $\pi$. If it is $0$ the skyrmions are bosons while if it is $\pi$ they are fermions. This field theory was once proposed\cite{polyakov} to describe the parent antiferromagnets of the cuprate materials. In the specific context of the square lattice Heisenberg antiferromagnet this proposal was killed by microscopic derivations of the sigma model\cite{nohopf} which revealed a Hopf coefficient of zero. With our modern understanding we can see that a Hopf coefficient of $\pi$ is not allowed in the presence of time reversal symmetry in any strictly $2d$ quantum magnet. Indeed this theory arises at the surface of the $3D$ topological superconductor. 

Our analysis of the $k=1$ topological superconductor implies that the non-linear sigma model with Hopf term at $\theta=\pi$ does not have a gapped phase that preserves the full $SO(3)\times\Z_2^T$ symmetry, even with topological order. This is an interesting conclusion that is not entirely obvious from other approaches.

As was seen in Sec.\ref{su(2)4cones}, the $k=2$ topological superconductor is also nontrivial under interaction. In particular, a gapped symmetric surface must necessarily develop topological order. Since the $k=2$ state can also be described using $4$ Dirac cones on the surface, the effective theory of the Neel order parameter $\vec{n}$ can be described using a non-linear sigma model with a Hopf term at $\theta=2\pi$. We therefore reach the surprising conclusion that, even a Hopf term at $\theta=2\pi$ cannot arise in a purely 2D system if time-reversal acts as $\vec{n}\to-\vec{n}$. Moreover, since the $k=4$ superconductor is trivial under interaction, a Hopf term with $\theta=4\pi$ is allowed in strict 2D with time-reversal.

\section{$U(1)\rtimes(\mathbb{Z}_2^T\times\mathbb{Z}_2^C)$: CII Class}
\label{CII}

Now we turn to fermions with charge $U(1)$, time-reversal and charge-conjugation symmetries that both square to $\T^2=\mathcal{C}^2=-1$ on physical fermions (the CII class). At free fermion level, the insulators are classified by $\Z_2$ in three dimensions. The non-trivial surface state has two Dirac cones:
\be
\label{c2}
H=\psi^{\dagger}(p_x\sigma_x+p_y\sigma_z)\otimes\tau_0\psi,
\ee
where the $U(1)$ symmetry acts in the obvious way, time reversal acts as
\be
\T: \psi\to i\sigma_y\otimes\tau_0\psi,
\ee
and charge conjugation acts as
\be
\mathcal{C}: \psi\to \sigma_0\otimes\tau_y\psi^{\dagger}.
\ee

The natural question to ask is how stable this phase is when interaction is included. Again we answer this question by looking at the $U(1)$ monopole. Notice that the composite operation $\mathcal{S}=\mathcal{T}\mathcal{C}$ is an anti-unitary operator that commutes with $U(1)$ rotation. Hence it plays the role of time-reversal in Sec.\ref{AIII}, where it was shown that the surface state with two Dirac cones gives a "Kramers" monopole. Therefore the monopole in the present case transforms as a Kramers pair under $\mathcal{S}$, which establishes the state as a nontrivial interacting SPT.

One may also ask that whether a (presumably strongly interacting) SPT exist for this symmetry group that gives a $\theta=\pi$ term in the $U(1)$ gauge response, since it looks consistent with symmetries but yet cannot be realized using free fermions. An analysis parallel to that in  Ref. \onlinecite{3dfSPT} and \onlinecite{fSTO2} shows that, however, such a state cannot exist. The basic idea is the following: if such a state exist, then combining the $(1,1/2)$ dyon (monopole carrying charge-$1/2$) with the $(-1,1/2)$ dyon gives the fundamental charge-$1$ fermion. A careful analysis then shows that $\mathcal{C}^2=1$ on such a composite, hence requiring the fundamental fermion to have $\mathcal{C}^2=1$ as well, which is inconsistent with the microscopic symmetry structure. Indeed, for microscopic symmetry such that $\mathcal{C}^2=1$, the state does exist, which is just the descendent of the electronic band TI with the additional symmetry $\mathcal{C}: \psi\to\psi^{\dagger}$.

Therefore the only non-trivial monopole structure is realized by the free fermion state Eq.~\eqref{c2}, which contributes a $\Z_2$ subgroup in the classification. The other SPT states, according to Sec.\ref{general}, are those from bosons with symmetry $\Z_2\times\Z_2^T$, which are classified\cite{chencoho2011,avts12} by $\Z_2^4$. The complete classification is thus given by $\Z_2^5$.

\section{$(U(1)\rtimes\mathbb{Z}_2^T)\times SU(2)$: $\mathbb{Z}_2^4$ classification from boson SPT}

Let us now turn to another physically relevant symmetry: charge conservation ($U(1)$), spin rotation ($SU(2)$) and time-reversal ($\T$). Free fermion band theory gives no non-trivial state, and we would like to examine it more carefully when interaction is included. Obviously one can always have SPTs coming from the charge-neutral bosonic sector, which have $SO(3)\times\Z_2^T$ symmetry, and are classified\cite{chencoho2011,avts12} by $\Z_2^4$. The real question is whether there is a strongly interacting SPT state not descending from bosonic sectors. According to Sec.\ref{general}, such states would necessarily have monopoles carrying nontrivial quantum numbers.

It is easy to first rule out a $\theta=\pi$ state\cite{3dfSPT,fSTO2}, where monopoles become charge-$1/2$ dyons: the bound state of the $(1,1/2)$ dyon and the $(-1,1/2)$ dyon, which are time-reversal partners, is the charge-$1$ physical fermion, which is an $SU(2)$-fundamental. It is then impossible to assign $SU(2)$ quantum numbers to either of the two dyons that is consistent with time-reversal symmetry.

Now we consider monopoles that are charge-neutral. The only nontrivial quantum number a monopole can carry is then an $SU(2)$-fundamental, since the $SU(2)$ group does not admit a projective representation. It turns out such a state does not exist as well, and the full classification is given simply by $\Z_2^4$ from bosonic SPT. We outline the argument briefly as follows:

We know that the monopole is bosonic and does not carry electric charge, so let's take advantage of that: instead of asking "could fermions give rise to spin-1/2 monopoles", let's ask the dual question instead: could spin-1/2 bosons give rise to fermionic monopoles? Now this becomes a question about boson SPT which is tractable, albeit with a less familiar symmetry.
Specifically the appropriate symmetry for these bosons is $U(1)\times SU(2)\times \Z_2^T$. Note that the contrast with the electrons (which are dual monopoles as seen by these bosons). 

The question can be further reduced to the following: does a boson SPT that gives a fermionic monopole survive if we further impose $SU(2)$ symmetry on the bosons, and require the charge-1 bosons transform as $SU(2)$ fundamental?

We then argue that for bosons with $U(1)\times \Z_2^T$, the SPT does not survive upon adding $SU(2)$ symmetry: for this symmetry group, $b_{\alpha} \to b_{\alpha}^{\dagger}$ under time-reversal, so the spin-up and down bosons do not get mixed under time-reversal. Therefore each spin-sector gives a time-reversal invariant boson insulator. More precisely, we can integrate out up-spin boson field since they are gapped anyway, the theory left behind contains only down-spin bosons, but it is still time-reversal invariant. Hence the two sectors should contribute equally to the $\theta$-angle in the $U(1)$ gauge response, which must be either $0$ or $2\pi$ due to time-reversal invariance in each sector. So the total $\theta$ must be $0$ or $4\pi$. It was shown in Ref. \onlinecite{avts12,hmodl,metlitski} that for boson systems $\theta=0$ and $4\pi$ correspond to a trivial insulator, while $\theta=2\pi$ gives an SPT state with fermionic monopoles (this is named as ''statistical Witten effect" in Ref. \onlinecite{metlitski}). Therefore it is impossible for the $U(1)\times SU(2)\times \Z_2^T$ bosons to induce a fermionic monopole.

The above argument does not work for bosons with $U(1)\rtimes \Z_2^T$, since a theory with only one species cannot be time-reversal invariant ($b_{\alpha}\to i\sigma_y^{\alpha\beta}b_{\beta}$ under $\T$), so each sector does not have to contribute to the $\theta$-angle in a time-reversal invariant way. For example, each sector can contribute a $\pi$ to $\theta$, so the total $\theta$ could be $2\pi$.

In the original (un-dual) problem, the above argument shows that it is impossible for fermions with $(U(1)\rtimes\Z_2^T)\times SU(2)$ symmetry to induce a monopole that transforms as $SU(2)$-fundamental. For fermions with $U(1)\times SU(2)\times \Z_2^T$ symmetry, on the other hand, it is possible for the monopole to carry spin-$1/2$ under $SU(2)$. In fact, it can be shown that the $k=1$ state discussed in Sec. \ref{su(2)2cones} survives upon imposing an extra $U(1)$ symmetry that commutes with $\T$, and the monopole of this $U(1)$ symmetry carries precisely spin-$1/2$ under $SU(2)$.

\section{Summary and discussion}
In this paper we studied the classification and physical properties of three dimensional interacting electronic topological insulators and superconductors. Free fermion systems in $3d$ fall into different symmetry classes described by the ``10-fold way". For all these  symmetry classes we were able to determine the stability to interactions, and further to determine if there are any new interacting phases that have no free fermion counterpart. If the symmetry group has a normal $U(1)$ subgroup we obtained the full classification in the presence of interactions. Our methods are physics-based and enable us to describe the physical properties of these various electronic SPT phases in three dimensions. 

We now discuss some open questions and some applications of our results. In the cases without a normal $U(1)$ subgroup it will be interesting to establish the completeness or lack thereof of our classification. For the symmetry groups $SU(2) \times Z_2^T$ or just $SU(2)$ in Section. \ref{c1}  we gave arguments why our classification may be complete. It is desirable to have a sharper version of these arguments. 

Perhaps the biggest open question about SPT phases is their possible occurrence in specific materials. For the 3d SPT phases with no free fermion counterpart for the most part we do not currently have  simple theoretical models which may be useful guides on the kinds of physical systems that are likely platforms for these phases. We hope that the enhanced understanding of these phases that our work provides will help answer such questions. 

An interesting application of our work, which we will elaborate elsewhere\cite{cwtsunpub}, is to the classification of three dimensional time reversal symmetric quantum spin liquids with an emergent photon (known as $U(1)$ spin liquids). These phases may be relevant to quantum spin ice materials. 
Though such quantum spin liquids are  ``long range entangled" they may nevertheless be fruitfully understood as gauged versions of SPT phases. The understanding of SPT phases provides a very insightful perspective on these time reversal symmetric quantum spin liquids. 

A different application of the results of this paper is widen the range of two dimensional quantum field theories which have anomalous implementation of symmetry. We showed that strictly two dimensions the non-linear sigma model description of collinear quantum antiferromagnets cannot have a Hopf term with a coefficient $\theta = \pi$ or $2\pi$. The former was proposed\cite{polyakov} as a possibility and discarded\cite{nohopf} on microscopic grounds. Our results show that Hopf terms with $\theta = \pi, 2\pi$ are consistent with time reversal only if the two dimensional magnet is the boundary of a three dimensional SPT phase. 

We thank A. Potter for discussions and a previous collaboration that was the stepping stone for this work. We also thank  Liang Fu for helpful discussions. This work was supported by NSF DMR-1305741.  This work was partially supported by a Simons Investigator award from the
Simons Foundation to Senthil Todadri.

\appendix
\section{Electric and thermal hall conductance mismatch}
\label{e8}

Here we discuss the constraints on quantum hall and thermal hall effect in a two-dimensional charged fermion system in the absence of intrinsic topological order and fractionalization. It is well known that in such cases the electric hall conductance $\sigma_{xy}$ is quantized in unit of $\sigma_0=e^2/h$, and the thermal hall conductance $\kappa_{xy}$ is also quantized in units of $\kappa_0=\frac{\pi^2}{3}\frac{k_B^2}{h} T$. For free fermions, the two should agree $\sigma_{xy}/\sigma_0=\kappa_{xy}/\kappa_0=n$ since the fermions transport both electricity and heat. 

With interactions, however, the two integers could differ. A simple example is the following: imagine and odd number (say $2n+1$) of fermions form a bound state $F\sim f^{2n+1}$, which is also a fermion but with charge $e^*=2n+1$. Now put the bound state fermion $F$ into a Chern band with Chern number $\nu$. The quantum hall conductance is then $\nu (e^*)^2=(2n+1)^2\nu$, but the thermal hall conductance is simply $\nu$ since it does not distinguish the charge carried by the fermion. The two quantized quantities thus have a mismatch 
\begin{eqnarray}
\label{mismatch}
\frac{\sigma_{xy}}{\sigma_0}-\frac{\kappa_{xy}}{\kappa_0}&=&((2n+1)^2-1)\nu \\ \nonumber
&=&8\left(\frac{n(n+1)}{2}\nu\right)=0 ({\rm mod} 8).
\end{eqnarray}

In general, it can be shown that the identity Eq.~\eqref{mismatch} is true as long as the system does not develop intrinsic topological order. We outline part of the proof here: for a system with any $\sigma_{xy}$ and $\kappa_{xy}$, we can stack it with certain integer quantum hall system made of free fermions so that the net $\sigma_{xy}$ becomes zero. If the remaining thermal hall conductance $\kappa'_{xy}/\kappa_0=\kappa_{xy}/\kappa_0-\sigma_{xy}/\sigma_0$ is non-zero, there must be a chiral edge mode that carries no charge (with chiral central charge $\kappa'_{xy}/\kappa_0$). But since the fermions are charged, the neutral charge mode must be bosonic. Hence it can be viewed as a boson state with a chiral edge. It is known that for a boson system with no topological order, $\kappa'_{xy}/\kappa_0=0({\rm mod} 8)$ (for a proof, see for example Ref. \onlinecite{luashvin}).

\section{Implication of a trivial monopole}
\label{tm}

Here we give the argument in Ref. \onlinecite{3dfSPT} for completeness. We will show that a trivial $U(1)$ monopole implies that the corresponding SPT phase must be equivalent to a boson SPT with no $U(1)$ charge.

It is convenient to start from a superconducting surface state which breaks the normal $U(1)$ subgroup but keeps the rest of the symmetries $G/U(1)$ unbroken. The suitable degrees of freedom then are $\frac{hc}{2e}$ vortices and (neutralized) Bogoliubov quasiparticles\cite{z2long}  (spinons) which have mutual semion interactions. In general it is possible for some exotic order to co-exist on the superconducting surface, such as intrinsic topological order, or even gapless degrees of freedom.

Now imagine tunneling a monopole from the vacuum to the system bulk. Since the monopole is trivial in both regions, the tunneling event - which leaves a $\frac{hc}{e}$ vortex   on the surface - also carries no non-trivial quantum number.  Hence the surface dual effective field theory has a bosonic $\frac{hc}{e}$-vortex that carries no non-trivial quantum number. We can therefore proliferate (condense) the $\frac{hc}{e}$-vortex on the surface  which disorders the superconductor and  
 yields an insulator  with the full symmetry $G$ unbroken.  However as is well known from dual vortex descriptions\cite{bfn,z2long} of spin-charge separation in $2D$, the resulting state has intrinsic topological order.  

In  this surface topologically-ordered symmetry-preserving insulator, a quasi-particle of charge-$q$ sees the $\frac{hc}{e}$-vortex as a $2\pi q/e$ flux.  Hence, the $\frac{hc}{e}$-vortex condensate confines all particles with fractional charge and quantizes the charge to $q=ne$ for all the remaining particles in the theory (for a more detailed discussion of this point, see Appendix C in Ref. \onlinecite{3dfSPT}). However, we can always remove integer charge from a particle without changing its topological sector by binding physical electrons. Hence the particle content of the surface topological order is $\{1,\epsilon,...\}\times\{1,c\}$, where only the physical electron $c$ in the theory is charged, and all the non-trivial fractional quasi-particles in $\{1,\epsilon,...\}$ are neutral. Since the $U(1)$ subgroup is normal, the action of $G/U(1)$ has to be closed within the neutral sector $\{1,\epsilon,...\}$. We can therefore describe the surface topological order as a purely charge-neutral quantum spin liquid with topological order $\{1,\epsilon,...\}$, supplemented by the presence of a trivial electron band insulator, $\{1,c\}$. In particular, any gauge-invariant local operator made out of the topological theory must be neutral (up to binding electrons), but in an electron system a local neutral object has to be bosonic. Hence the theory should be viewed as emerging purely from a neutral boson system. This implies that the bulk SPT order should also be attributed to the neutral boson sector, with the symmetry $G/U(1)$.

\end{document}


\title{Erratum: interacting fermionic topological insulators/superconductors in three dimensions}
\author{Chong Wang and T. Senthil}
\affiliation{Department of Physics, Massachusetts Institute of Technology,
Cambridge, MA 02139, USA}
\date{\today}

\maketitle

We correct an error in our paper Ref.~\onlinecite{ws14}. It was claimed in Ref.~\onlinecite{ws14} that interacting topological insulators of non-Kramers fermions ($\mathcal{T}^2=1$) -- also known as the AI class in the literature -- were classified by $\mathbb{Z}_2^2$. Here we show that the correct classification should be $\mathbb{Z}_2$. This phenomenon is unique to non-Kramers insulators (AI class) in the periodic table, and all the other classification results in Table.~I of Ref.~\onlinecite{ws14} are unchanged. We summarize the modified results here in Table~\ref{classtable}.

 \begin{table}[h]
\begin{center}
\begin{tabular}{|>{\centering\arraybackslash}m{1.1in}|>{\centering\arraybackslash}m{0.7in}|>{\centering\arraybackslash}m{0.6in}|>{\centering\arraybackslash}m{0.8in}|}
\hline
{\bf Symmetry class} &  {\bf Reduction of free fermion states} & {\bf Distinct boson SPT} & {\bf Complete classification} \\ \hline
$U(1)$ only (A) & $0$ & $0$ & $0$ \\ \hline
$U(1)\rtimes\Z_2^T$, $\T^2=-1$ (AII) & $\Z_2\to\Z_2$ & $\Z_2^2$ & $\Z_2^3$ \\ \hline
$U(1)\rtimes\Z_2^T$, $\T^2=1$ (AI) & $0$ & \color{red}{$\Z_2$} & \color{red}{$\Z_2$} \\ \hline
\hline
$U(1)\times\Z_2^T$ (AIII) & $\Z\to\Z_8$ & $\Z_2$ & $\Z_8\times\Z_2$ \\ \hline
$U(1)\rtimes(\Z_2^T\times\Z_2^C)$ (CII) &  $\Z_2\to\Z_2$ & $\Z_2^4$ & $\Z_2^5$ \\ \hline
$(U(1)\rtimes\Z_2^{T})\times SU(2)$ & 0 & $\Z_2^4$ & $\Z_2^4$ \\ \hline
\hline
$\Z_2^T$, $\T^2=-1$ (DIII) & $\Z\to\Z_{16}$ & $0$ & $Z_{16}$ (?) \\ \hline
$SU(2)\times\Z_2^T$ (CI) & $\Z\to\Z_4$ & $\Z_2$ & $Z_4 \times Z_2$ (?) \\ \hline

\end{tabular}
\end{center}
\caption{Summary of results on classifications of fermionic SPT states in three dimensions. Results for the AI class (marked red) are modified. The rest remain the same as Table~I in Ref.~\onlinecite{ws14}. } 
\label{classtable}
\end{table}%

The key is to show that the topological paramagnet $eTmT$ -- a bosonic symmetry-protected topological (SPT) phase protected by time-reversal symmetry -- becomes trivial when the system contains physical fermions that are charged and non-Kramers ($\mathcal{T}^2=1$).
To see this, we consider the surface topological order of the $eTmT$ state, which is a $\mathbb{Z}_2$ topological order, with quasi-particles labeled as $\{1,e,m,\epsilon\}$ and both $e$ and $m$ have $\mathcal{T}^2=-1$. We now show that in the presence of physical non-Kramers fermions, this $eTmT$ topological order becomes realizable in strict two dimensions. Therefore the surface topological order can be trivialized (confined) without breaking the $U(1)\rtimes\T$ symmetry, and the corresponding bulk state should also be trivial.

In the presence of non-Kramers local fermions $f$, one can combine the fermion with the $e$ and $\epsilon$ particle in the $eTmT$ topological order. This is essentially a relabeling of the same phase. The resulting $\mathbb{Z}_2$ topological order has $\tilde{e}=f\epsilon$ that has charge-$1$ but is non-Kramers, and $\tilde{m}=m$ that is Kramers but charge-neutral. This topological order, dubbed $eCmT$ in Ref.~\onlinecite{hmodl}, turns out to be realizable even in strictly two dimensional systems. Hence it is anomaly-free. To realize this state in strict $2D$, one can start from the $eCT\epsilon CT$ state -- a state in which both $e$ and $\epsilon$ carry charge-$1$ and are Kramers ($\mathcal{T}^2=-1$). The $eCT\epsilon CT$ state is anomaly-free, because the $m$ particle carries no nontrivial quantum number, and can be condensed to confine the topological order without breaking any symmetry. Then one can put the $\epsilon$ particle into a $2D$ topological insulating band. The resulting state is well known\
cite{eCmT} to be the 
$eCmT$.